\documentclass[reqno,12pt]{amsart} 
\usepackage[top=1.5in,right=1.125in,left=1.125in,bottom=1.5in]{geometry}
\usepackage{amssymb}
\usepackage{color}
\usepackage{tikz}
\usetikzlibrary{positioning,decorations.pathmorphing}
\usetikzlibrary{positioning,decorations.pathmorphing}
\definecolor{red}{rgb}{0.7,0,0}
\definecolor{grey}{RGB}{112,112,112}
\definecolor{blue}{RGB}{034,113,179}
\usepackage[colorlinks=true,citecolor=blue,linkcolor=red,urlcolor=grey,backref=page]{hyperref}

\theoremstyle{remark}

\newcounter{mnotecount}[section]

\renewcommand{\themnotecount}{\thesection.\arabic{mnotecount}}

\newcommand{\mnote}[1]
{\protect{\stepcounter{mnotecount}}$^{\mbox{\footnotesize
$
\bullet$\themnotecount}}$ \marginpar{
\raggedright\tiny\em
$\!\!\!\!\!\!\,\bullet$\themnotecount: #1} }

\newcommand{\HH}{\mathbb{H}}

\newcommand{\R}{\mathbb{R}}

\def\p{\partial}
\def\be{\begin{equation}}

\def\ee{\end{equation}}

\def\bea{\begin{eqnarray}}
\def\eea{\end{eqnarray}}

\numberwithin{equation}{section}
\begin{document} \date{September 25th, 2021}
\title{The Kijowski--Liu--Yau quasi-local mass of the Kerr black hole horizon}
\author{Maciej Dunajski}
\address{Department of Applied Mathematics and Theoretical Physics\\ 
University of Cambridge\\ Wilberforce Road, Cambridge CB3 0WA, UK.}
\email{m.dunajski@damtp.cam.ac.uk}
\author{Paul Tod}
\address{The Mathematical Institute\\
Oxford University\\
Woodstock Road, Oxford OX2 6GG\\ UK.
}
\email{tod@maths.ox.ac.uk}
\maketitle
\begin{center}
{\em Dedicated to Gary Gibbons on the occasion of his 75th birthday.}
\end{center}
\begin{abstract}
  We use an isometric embedding of the
cross-over surface of the outer
  horizon of a rapidly rotating Kerr black hole in a hyperbolic space  to compute the quasi-local mass of the horizon for any allowed value of the spin parameter $j=J/m^2$. The mass is monotonically decreasing
from twice the ADM mass at $j=0$ to $1.76569m$ at $j=\sqrt{3}/2$. It then monotonically increases to a maximum around
$j=0.99907$, and finally decreases to $2.01966m$ for $j=1$ which corresponds to the extreme Kerr black hole.
\end{abstract}
\section{Introduction}
There is no local concept of energy density in general relativity, which makes
defining the mass of a local system problematic. On the other hand
the ADM and the Trautman--Bondi masses 
of asymptoticaly flat space times  can be defined at, respectively, spacelike and null infinities, and their positivity can be established given appropriate
energy and regularity conditions.
A quasi-local energy aims be a compromise, as it
associates a number to  any closed space--like two--surface $\Sigma$ in space time. There are several approaches to the subject which are reviewed in \cite{szabados}. In particular
the prescriptions of Kijowski \cite{K}
and Liu--Yau \cite{LY}, which are related to a class of definitions proposed by Brown and York \cite{BY},  are applicable to surfaces
with non--negative Gaussian curvature. This condition guarantees, by
Nirenberg's solution \cite{Ni} of the Weyl embedding problem, the existence of a unique (up to an isometry of $\R^3$) global isometric embedding of $\Sigma$ in $\R^3$.
The resulting mass is then defined, up to a constant factor, to be an integral of the difference between the 
mean curvature of this flat embedding, and the norm of the  mean curvature vector of $\Sigma$ regarded as a surface in the space--time. This prescription is therefore not applicable to a space--like section of 
the horizon of rapidly--rotating Kerr black hole. This is because if the dimensionless spin parameter $j\equiv J/m^2$ (where $J$ is the angular momentum, and $m$ is the ADM mass) is between $\sqrt{3}/2$ and $1$, then the Gaussian curvature  is negative near north and south poles. It is this problem which we resolve in this note.
It will be done by embedding the Kerr horizons with $j>\sqrt{3}/2$ in a hyperbolic space and choosing  the hyperbolic radius so that the embedding continuously
matches with that in $\R^3$ for $j=\sqrt{3}/2$.

In \S\ref{section2} we shall compute the mean curvatures of both the
hyperbolic and flat embeddings, and
in \S\ref{section3} we shall construct the corresponding quasi--local
mass as a function of $j\in [0, 1]$. This function is monotonically decreasing
from twice the ADM mass $m$ at $j=0$ to  $1.76569m$ at $j=\sqrt{3}/2$.
The energy then  monotonically increases to the value of $2.02223m$ reached around
$j=0.99907$, and decreases  down to
$2.01966m$ which is the quasi local mass of the extreme Kerr 
horizon where $j=1$.
\subsection*{Acknowledgements} 
MD has been partially supported 
by STFC grants ST/P000681/1, and  ST/T000694/1. We are grateful to Adam Dunajski, Christian Klein, and
especially Ulrich Sperhake for their help with numerical integration, and to Don Page for the correspondence about Kerr black holes with values of $j$ close to $1$.

\section{Isometric embeddings of Kerr black hole horizons}
\label{section2}
\subsection{Hyperbolic embedding of rapidly rotating horizons}
Let $(\Sigma, g)$ be a two--dimensional Riemannian manifold with the Gaussian curvature bounded below by
a negative constant $-L^{-2}$. The Pogorelov theorem \cite{P} states that there is a global isometric
embedding of $\Sigma$ into hyperbolic three--space $\HH^3$ with
Ricci scalar less than or equal to $-6L^{-2}$.
We shall consider the upper half--space model of $\HH^3$, with the hyperbolic metric
\be
\label{GH3}
G_L=\frac{L^2}{z^2}\Big(dz^2+dr^2+r^2d\phi^2\Big), \quad z>0, r>0, \phi\in [0, 2\pi).
\ee
In \cite{G} a global isometric embedding of $(\Sigma, g)$ was explicitly constructed in the case where
$g$ admits a $U(1)$ isometric action, and such that this isometry preserves the embedding. This result was then used in \cite{G} to construct an isometric embedding of the spatial sections of Kerr black hole horizons.
We shall first reproduce this embedding, and then compute its extrinsic properties: the second fundamental form, and the mean curvature. 
 
 Consider an embedding 
$\iota:\Sigma \rightarrow \HH^3$,
where $(\Sigma, g)$ is a surface of revolution with coordinates
$x\in[-1, 1], \phi\in [0, 2\pi)$
and
\be
\label{kerr1}
g=\rho^2(B^{-1} dx^2+Bd\phi^2), \quad B=B(x),\quad \rho=\mbox{const}.
\ee
If $z=Z(x), r=R(x)$, then
$g=\iota^*(G_L)$ iff
\begin{eqnarray}
\label{Z}
Z(x)&=&\exp{\Big(\int\Big(\frac{-\rho^2 BB'\pm \rho\sqrt{B(4\rho^2B+4L^2-L^2(B')^2)}}{2B(\rho^2B+L^2)}\Big)dx\Big)},\\
R(x)&=&\frac{\rho}{L}\sqrt{B(x)}Z(x).\nonumber
\end{eqnarray}
The mean curvature $H$ of this embedding with respect to the outward pointing
unit normal vector field $N$ to $\Sigma$ in $\HH^3$, where
$
N=\frac{Z}{L\sqrt{(R')^2+(Z')^2}}\Big(R'\frac{\p}{\p Z}-Z'\frac{\p}{\p R}\Big)
$,
is a half of the $g$--trace of the second fundamental form
$h$. Using the definition 
$
h(X, Y)=G_L(N, \nabla_X Y),
$
where $(X, Y)$ are the elements of $T\Sigma$, 
and $\nabla$ is the Levi--Civita connection of
$G_L$ we find
\be
\label{mean_first}
H=\frac{-RZZ'R''+RZR'Z''+2((R')^2+(Z')^2)(RR'+ZZ'/2)}{2LR((R')^2+(Z')^2)^{3/2}}.
\ee
We aim
to match (\ref{kerr1})  with the general form of the  Kerr horizon metric
with the ADM mass $m$, and the angular momentum $0\leq J\leq m^2$.
To do it, consider the Kerr metric written in the Boyer–-Lindquist coordinates (see, e.g. \cite{wald}), and
restrict it to a surface of constant time on the outer event horizon, which gives
\be
\label{kerrBL}
g=Sd\theta^2+\Big({r_+}^2+\frac{J^2}{m^2}+\frac{2J^2r_+}{mS}\sin^2{\theta}\Big)\sin^2{\theta}d\phi^2,
\ee
where
\[
r_+=m+\sqrt{m^2-J^2/m^2}, \quad S={r_+}^2+(J^2/m^2)\cos^2{\theta}.
\]
The metric (\ref{kerr1}) then arises from (\ref{kerrBL}) by setting $x=\cos{\theta}$, adopting $(x, \phi)$ as coordinates, and taking
\[
B=\frac{(1+c^2)(1-x^2)}{1+c^2x^2},\quad
\rho^2=2m(m+\sqrt{m^2-J^2/m^2}), \quad c=\frac{2J}{\rho^2}\in[0, 1], \quad x\in [-1, 1].
\]
The constant $\rho$ is twice the irreducible mass of the Kerr black hole (so that it is also proportional to the square root of the area of the outer horizon), and we choose  a plus  sign in (\ref{Z}).

The Gaussian curvature $K$ of $g$ is bounded from below
\be
\label{Gaussian_cur}
K=\frac{(c^2+1)^2(1-3c^2x^2)}{\rho^2(1+c^2x^2)^2}\geq \frac{(1-3c^2)}{\rho^2(c^2+1)}=K_{min}.
\ee
For $c\in(\sqrt{3}^{-1}, 1]$ we take $L=\rho\sqrt{\frac{1+c^2}{3c^2-1}}$, which is the largest hyperbolic radius for which the embedding is global. This, after some calculations, gives the mean curvature (\ref{mean_first}) as
\be
\label{Hc}
H=\frac{c\sqrt{1-x^2}(x^4(2c^4-5c^6)+x^2(-4c^6-12c^4+6c^2)+c^4+3c^2+9)}
{\rho\sqrt{(1+c^2)(1+c^2x^2)^3}\sqrt{x^4(c^4-2c^6)+ x^2(-c^6-4c^4+3c^2)+3}}.
\ee
\subsection{Flat embedding of slowly rotating horizons}
The hyperbolic embedding (\ref{Z}) with the critical choice of the hyperbolic radius is well defined as long as
$\sqrt{3}^{-1}< c\leq 1$, which corresponds to the spin parameter $j\in [\sqrt{3}/2, 1]$. If $c\leq \sqrt{3}^{-1}$ then the Kerr horizon can be globally isometrically embedded in $\R^3$. If the flat metric on $\R^3$ is
\[
G=d\zeta^2+dr^2+r^2d\phi^2,
\]
then the embedding is given by
\be
\label{zeta}
\zeta=\pm\frac{\rho}{2}\int\frac{\sqrt{B(4-(B')^2)}}{B}dx, \quad
r=\rho\sqrt{B}.
\ee
The formulae (\ref{zeta}) can also be obtained as a limiting case of the hyperbolic embedding when the hyperbolic radius tends to infinity.
To see this, set $
z=Le^{\zeta/L}, L>0
$
and find
\[
G=\lim_{L\rightarrow\infty} G_L, \quad\mbox{where}\quad 
G_L= d\zeta^2+e^{-2\zeta/L}(dr^2+r^2d\phi^2).
\]
For (\ref{zeta}) to be well defined we need
$B(4-(B')^2)\geq 0$ which gives
\[
c^8(x^6+x^4+x^2)+4c^6(x^4+x^2)+6c^4x^2-1\leq 0
\]
which should hold for all $x\in[0, 1]$. Evaluating the expression above at
$x^2=1$ gives a polynomial with two real 
roots $c=\pm\sqrt{3}/3$,
and the inequality holds when $0\leq c\leq 1/\sqrt{3}$. This is the same condition
which guarantees $B''\leq 0$, which holds iff the Gaussian curvature is non--negative. Thus, although $K\geq 0$ is necessary and sufficient for
(\ref{zeta}) to be a global isometric embedding, there can be regions
on the Kerr horizon where the Gaussian curvature is negative, and yet the embedding still exists (although it does not extend to the whole horizon).

Repeating the steps leading to (\ref{Hc}) we find that the mean curvature of the embedding (\ref{zeta}) is given by
\[
H_0=-\frac{c^8x^6+(c^8+4c^6)x^4+(4c^8+13c^6+15c^4+3c^2)x^2-(c^6+3c^4+2c^2+2)}{2\rho(c^2x^2+1)\sqrt{(1+c^2)(1+c^2x^2)(1-(x^3+x^2+x)c^4-2c^2x)(1+ (x^3+x^2+x)c^4+2c^2x)}}.
\]
\begin{center}
\label{fig1}
\includegraphics[width=5cm,height=5cm,angle=0]{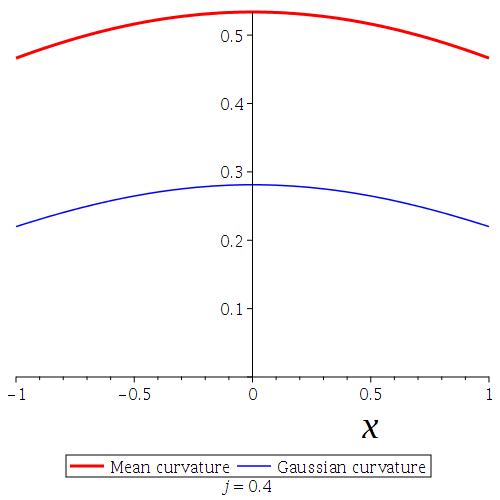},
\includegraphics[width=5cm,height=5cm,angle=0]{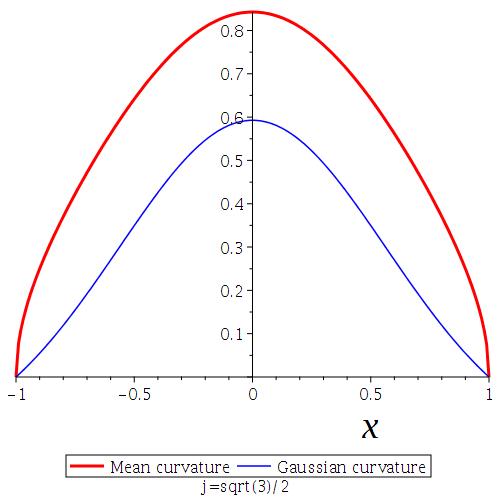},
\includegraphics[width=5cm,height=5cm,angle=0]{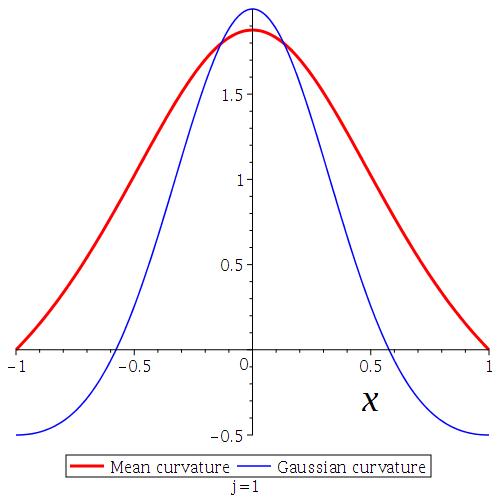}
\end{center}
{{\em  {\bf Figure 1.} Mean (in red) and Gaussian (in blue) curvatures of the Kerr horizons for various values of the spin parameter $j$.}}
\vskip5pt
To sum up, the Kerr horizon with any $c\in [0, 1]$ can be globally and isometrically embedded as a hypersurface in the space of constant
curvature (Figure 1). If $c\leq 1/\sqrt{3}$ then the embedding is in $\R^3$ and when $c>1/\sqrt{3}$ then it is in $\HH^3$. The mean curvatures of the flat and hyperbolic embeddings are equal at $c=1/\sqrt{3}$, but 
their derivatives w.r.t. $c$ are different. This will play a role
in analysing the behaviour of the quasi-local energy as a function of $c$. We shall do this in the next section.
\section{Modified Kijowski--Liu--Yau Mass}
\label{section3}
The Kijowski--Liu--Yau \cite{K, LY} definition of quasi--local mass of a space--like closed 2-surface
$\Sigma$ in a space--time $M$
is 
\be
\label{KLY}
E_{KLY}=\frac{1}{4\pi}\int_{\Sigma} (H-|\hat{H}|) \mbox{vol}_{\Sigma},
\ee
where $H$ is the mean curvature of the embedding of $\Sigma$ in $\R^3$, and $|\hat{H}|$ is the space--time
norm of the mean curvature vector $\hat{H}$ of the surface $\Sigma$ embedded in $M$. This is well defined only if $\hat{H}$ is space--like or zero, and if the Gaussian curvature of 
$\Sigma$ is non--negative, as then
a global embedding of $\Sigma$ in $\R^3$ exists.
\vskip5pt
In what follows, we modify the KLY definition replacing the embedding in $\R^3$ by 
the embedding in $\HH^3$, where the hyperbolic radius $L$ of $\HH^3$ is maximal for which the embedding exists, i.e. 
such that $K_{min}=-L^2$, where $K_{min}$ given by (\ref{Gaussian_cur}) is the lower bound for the Gaussian curvature of $\Sigma$.
If $\Sigma$ is taken to be the surface of the horizon,
then $\hat{H}=0$, and the mass is proportional to the integral of the mean curvature. This modification of the KLY mass also leads to a non--negative
expression (Theorem 3.1 in \cite{YW0}).

Set $j=J/m^2$. We first restrict the range of $j$ 
to $(\sqrt{3}/2, 1]$, which corresponds to the hyperbolic radius
between $\infty$ and $\sqrt{2}m$, the latter case corresponding to the extremal Kerr metric, and the former case corresponding to $K_{min}=0$. For the mean curvature (\ref{Hc}) we compute the modified mass to be
\begin{eqnarray}
\label{mass_final}
E(m, j)&=& \frac{1}{4\pi}\int_{-1}^{1}\int_0^{2\pi}  H(x) \rho^2 d\phi d x\nonumber\\
&=&
m\int_{-1}^1
\frac{\sqrt{1-x^2}(x^4(2c^5-5c^7)+x^2(-4c^7-12c^5+6c^3)+c^5+3c^3+9c)}
{2{(1+c^2)(1+c^2x^2)^{3/2}}\sqrt{x^4(c^4-2c^6)+ x^2(-c^6-4c^4+3c^2)+3}}dx\nonumber\\
&&\mbox{where}\quad c=\frac{1-\sqrt{1-j^2}}{j}, \quad \rho=\frac{2m}{\sqrt{1+c^2}}.
\end{eqnarray}
The limiting values of the mass are
\begin{eqnarray}
\label{mass32}
E(m, \sqrt{3}/2)&=&m\int_{-1}^{1}\frac{3\sqrt{1-x^2}(x^4+14x^2+273)}{8(x^2+3)^{3/2}\sqrt{3x^4+42x^2+243}}dx\approx 1.76569m,\\
E(m, 1)&=&m\int_{-1}^1 \frac{13-10x^2-3x^4}{\sqrt{(x^2+3)(x^2+1)^3}}dx\approx 2.01966m.
\end{eqnarray}
Using the flat embedding in the range $c\in [0, \sqrt{3}^{-1}]$ 
we find
\be
E(m, j)=
m\int_{-1}^1\frac{-c^8x^6-(c^8+4c^6)x^4+(4c^8+13c^6+15c^4+3c^2)x^2+c^6+3c^4+3c^2+2}{2(c^2x^2+1)^{3/2}(c^2+1)^{3/2}\sqrt{(1-c^4(x^3+x^2+x)-2c^2x)(1+c^4(x^3-x^2+x)+2c^2x)} }dx.
\ee
The limiting value at $j=\sqrt{3}/2$ agrees with (\ref{mass32}). The other limit is $E(m, 0)=2m$.

The mean curvature as a function of the spin parameter $j=J/m^2$
is continuous but not smooth at $j=\sqrt{3}/{2}$ which separates the flat and the hyperbolic embeddings. For $0\leq j\leq \sqrt{3}/2$ the
quasi local mass is a decreasing function of $j$, and is very well approximated by the first four terms of the series
\be
\label{slow_in}
E(m, j)= m\Big(2-\frac{1}{4}j^2-\frac{17}{320}j^4-\frac{407}{17920}j^6-
\dots  \Big).
\ee
In the Schwarzschild case $j=0$ the quasi-local energy is equal to twice the ADM mass in agreement with the results of Martinez \cite{Martinez}, who (unlike us)
additionally assumed that $j\ll 1$, and only derived the first two terms
in the series (\ref{slow_in}). Our findings also  disprove
the conjecture of Martinez, that the quasi-local energy is equal to twice the
irreducible mass. Expanding the latter quantity (which is equal to our $\rho$)
we find
\[
\rho= m\Big(2-\frac{1}{4}j^2-\frac{25}{320}j^4-\frac{735}{17920}j^6-
\dots  \Big)<E(m, j)\quad\mbox{if}\quad j\in \Big(0, \frac{\sqrt{3}}{2}\Big].
\]
Thus, in this range of $j$, the quasi--local mass is always greater
than twice the irreducible mass (Figure 2) which is in agreement
with the Minkowski inequality
\[
\frac{1}{4\pi}\int_{\Sigma} H\mbox{vol}_{\Sigma} \geq \frac{1}{4\pi}\sqrt{4\pi\mbox{Area}(\Sigma)}=\rho.
\]
\begin{center}
\label{fig2}
\includegraphics[width=5cm,height=5cm,angle=0]{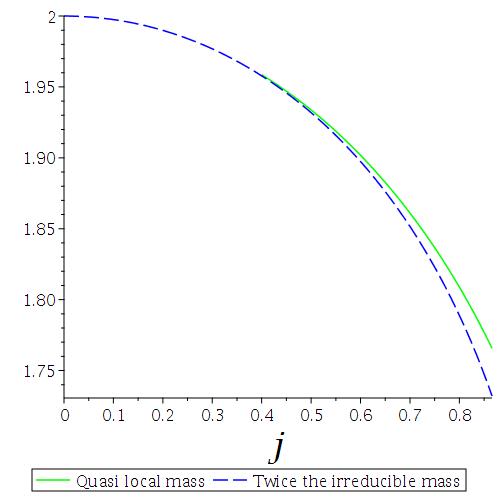}.
\end{center}
    {{\em  {\bf Figure 2.} The quasi local mass is greater than twice
        the irreducible mass.
}}
    \vskip2pt
    
As the spin parameter increases to $\sqrt{3}/2$, the energy decreases
to $1.76569m$. For $j$ above this value, the original Brown--York--Kijowski--Liu--Yau prescription breaks down as the global isometric embedding in $\R^3$ does not exist. This was noted by Martinez, who states in \cite{Martinez} that his calculations are applicable only to slowly spinning Kerr black holes.
In the range $j\in (\sqrt{3}/2, 1]$ we use the hyperbolic embedding. This has a free parameter - the hyperbolic radius $L$ - constrained by the inequality
$
0<L\leq\frac{2m}{\sqrt{3c^2-1}},
$
and, for each $c$ we choose the maximal value of $L$ which makes the embedding global.
We then numerically integrate (\ref{mass_final}) to compute the energy as a function of $j$.
We have used the Clenshaw--Curtis quadrature method implemented on MAPLE 2020, and have verified that increasing the precision, and at the same adding more digits to $j$ does not change the first six digits in $E(j)$.  We have independently verified the result using a Python code 
and, applying  Simpson's rule, which we tested for convergence. We found the convergence between 1st and 2nd order, and so significantly lower than the 4th order
expected from the method. This slow convergence is caused by the presence of square roots in the integrand which makes the function not differentiable at the boundaries. Indeed, repeating the test, but instead integrating between $\pm 0.9$ yields a convergence factor of $15.4$ which is close enough to $16$
expected from 4th order Simpson's method. The numerical value of energy appears to be very precise for all values of $j$. Changing the resolution between $500$ and 
$10000$ points does not change the leading six digits.

 The energy increases in an almost linear way (a closer analysis shows that the graph deviates from a line). Zooming near $j=1$ shows that the energy reaches a maximum
of $2.02223m$ around $j=0.99907$, and then decreases to $2.01966m$ for the extremal Kerr horizon corresponding to $j=1$ (Figure
3). To show that this maximum
is not a numerical artefact we can expand the integrand in (\ref{mass_final}) near $j=1$, and find the integral to the lowest order in $(j-1)$:
\[
E(j)\approx E(1)+\sqrt{1-j}\;0.17459m
\]
so that $E(j)$ is indeed a decreasing function of $j$ near $j=1$.
\begin{center}
  \label{fig22}
\includegraphics[width=5cm,height=5cm,angle=0]{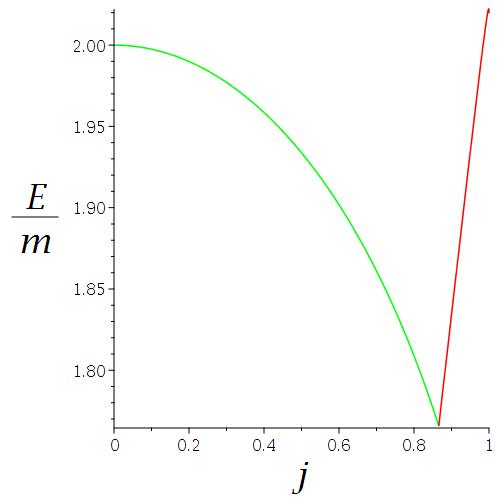},
\includegraphics[width=5cm,height=5cm,angle=0]{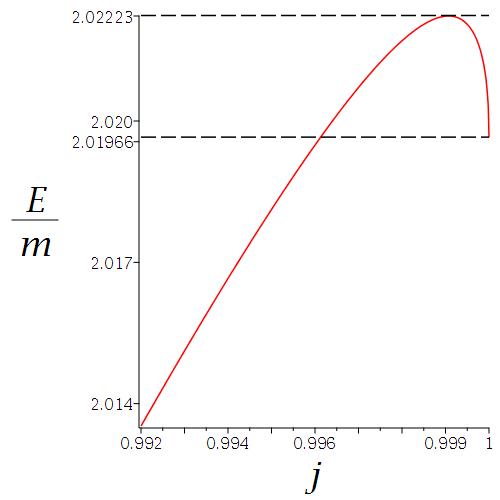}.
 \end{center}
    {{\em  {\bf Figure 3.} The quasi--local energy as a function of the spin parameter $j=J^2/m$. 
}}
\vskip2pt

In our computation of $E_{KLY}$ from the hyperbolic embedding, we have
chosen the hyperbolic radius of the ambient $\HH^3$ to be maximal such 
the embedding is global. This choice has ensured the continuity of the mean curvature as well as the resulting energy at $j=\sqrt{3}/2$.
We could instead leave $L$ as a positive parameter, and regard  (for each value
of $j$) the energy as a function of $L$. It turns out that this function is monotonically decreasing from $L=0$ to the critical value (Figure 4), which again suggests that our initial choice was a right one. 
\begin{center}
\label{fig3}
\includegraphics[width=5cm,height=5cm,angle=0]{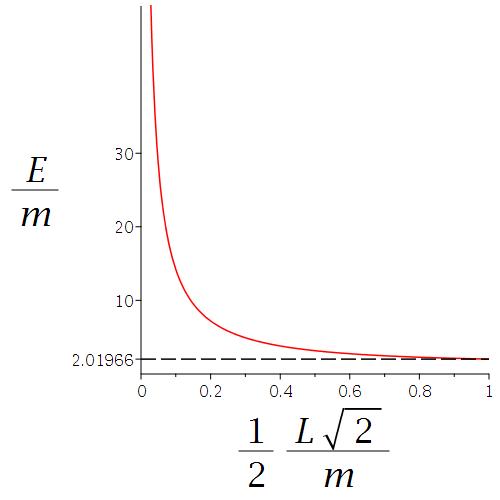},
\end{center}
{{\em  {\bf Figure 4.} The quasi-local energy of the extreme Kerr horizon
computed using the hyperbolic embedding as a function of the hyperbolic radius.
}}
\section{Conclusions} We have used a combination of flat and hyperbolic
embeddings to compute the quasi--local mass of the Kerr black hole
horizon for any allowed value of the angular momentum. The hyperbolic embedding
of the rapidly rotating horizons allowed us to overcome the difficulty arising
from the Gaussian curvature not being positive everywhere on $\Sigma$.
There is another, by now well established, way around this problem due to
Wang and Yau \cite{YW, YW2}, who used an embedding 
of $\Sigma$ in  $\R^{3, 1}$ to construct a reference frame. As well as allowing
for non--positive Gaussian curvature, the Wang--Yau approach addresses
a problem (pointed out in  \cite{MST}) that $E_{KLY}$ is positive on some
closed 2-surfaces in Minkowski space which do not lie in a space--like hyper--plane. The computation of the Wang--Yau quasi local
mass involves taking an infimum of all mean curvature integrals over all `time functions' $\tau$
such that the Gaussian curvature of $\hat{g}=g+d\tau^2$ is positive. It is therefore difficult to implement for concrete examples. This has nevertheless been attempted
in \cite{yau_kerr}, where the authors noted the difficulty of
finding a non--zero admissible time function as in general they lead to complex energies. By examining the boundary separating the complex and real energies
they have confirmed the result of Martinez \cite{Martinez} in the range $[0, 0.4]$, and improved it up to
$j\leq{\sqrt{3}/2}$. For $j\in(\sqrt{3}/2, 1]$ the numerical computations of \cite{yau_kerr} suggest that the mass is increasing which agrees with our findings. The analysis of \cite{yau_kerr} has not however revealed the global maximum\footnote{While we can not offer any physical explanation of this maximum, there exist at least two other occurrences of near extremal, but not extremal values of $j$ in astrophysics, and General Relativity:
the value $j\approx 0.998$ is needed for the equilibrium of a black hole absorbing matter
and radiation from an accretion disk \cite{Thorne}. In the context of
of rotating photon orbits in the Kerr solution it  has been shown that  
if $j\approx 0.99434$, then a photon sent out in a constant radial direction from the north polar axis returns to the north polar axis in the opposite direction. An effect
which Page calls `a photon boomerang' \cite{Page}.} of mass just before $j=1$. Additionally, the Wang--Yau mass is not defined
  if the mean curvature vector of $\Sigma$ in the space--time vanishes, which is the case for the cross--over surface of the outer horizon. In \cite{yau_kerr}
the mass was therefore calculated at a constant radius, and the outer-horizon limit was taken. It is however not clear whether this limit is unique. 

\end{document}